\documentclass[conference]{IEEEtran}
\IEEEoverridecommandlockouts
\usepackage{cite}
\usepackage{amsmath,amssymb,amsfonts}
\usepackage{algorithmic}
\usepackage{graphicx}
\usepackage{textcomp}
\usepackage{xcolor}
\usepackage{multirow}
\def\BibTeX{{\rm B\kern-.05em{\sc i\kern-.025em b}\kern-.08em
 T\kern-.1667em\lower.7ex\hbox{E}\kern-.125emX}}
\usepackage[normalem]{ulem}
\def\KeepNewOnly{1}
\def\EnabledValue{1}
\newcommand\camerareadychanges[2]{%
\ifx\KeepNewOnly\EnabledValue%
    #2%
\else%
    \sout{\textcolor{blue}{#1}}\textcolor{red}{#2}%
\fi}
\begin{document}

\title{More Similar than Dissimilar: Modeling Annotators for Cross-Corpus Speech Emotion Recognition
% \thanks{Identify applicable funding agency here. If none, delete this.}
}

% \author{\IEEEauthorblockN{Anonymous Submission}
% \IEEEauthorblockA{Anonymous Submission \\
% Anonymous Submission\\
% Anonymous Submission \\
% Anonymous Submission}

\author{\IEEEauthorblockN{1\textsuperscript{st} James Tavernor}
\IEEEauthorblockA{\textit{Computer Science and Engineering} \\
\textit{University of Michigan}\\
Ann Arbor, United States \\
tavernor@umich.edu}
\and
\IEEEauthorblockN{2\textsuperscript{nd} Emily Mower Provost}
\IEEEauthorblockA{\textit{Computer Science and Engineering} \\
\textit{University of Michigan}\\
Ann Arbor, United States \\
emilykmp@umich.edu}
% \and
% \IEEEauthorblockN{3\textsuperscript{rd} Given Name Surname}
% \IEEEauthorblockA{\textit{dept. name of organization (of Aff.)} \\
% \textit{name of organization (of Aff.)}\\
% City, Country \\
% email address or ORCID}
% \and
% \IEEEauthorblockN{4\textsuperscript{th} Given Name Surname}
% \IEEEauthorblockA{\textit{dept. name of organization (of Aff.)} \\
% \textit{name of organization (of Aff.)}\\
% City, Country \\
% email address or ORCID}
% \and
% \IEEEauthorblockN{5\textsuperscript{th} Given Name Surname}
% \IEEEauthorblockA{\textit{dept. name of organization (of Aff.)} \\
% \textit{name of organization (of Aff.)}\\
% City, Country \\
% email address or ORCID}
% \and
% \IEEEauthorblockN{6\textsuperscript{th} Given Name Surname}
% \IEEEauthorblockA{\textit{dept. name of organization (of Aff.)} \\
% \textit{name of organization (of Aff.)}\\
% City, Country \\
% email address or ORCID}
}

\maketitle
\begin{abstract}
% EP note: I've made changes
% Speech emotion recognition (SER) systems are often trained to predict an aggregate label, typically the average of a group of annotators. The benefit to this approach is that it mitigates the variability inherent in the annotations of any one person. The drawback is that these models struggle to predict how any one person would perceive an emotional display. The alternative is to train individual prediction models, or prediction heads, for each annotator. Yet, extending these models to new annotators requires data and computational resources to train new models. However, annotators are often similar to each other. Therefore, we propose that it is possible to approximate models personalized to a new annotator by deploying a ``similar'' pre-trained annotator-specific model in a zero-shot manner. The benefit is that this would not require retraining and instead would require only a small amount of enrollment data for the new annotators. We demonstrate that our approach significantly outperforms other zero-shot approaches, paving the way for lightweight emotion adaptation practical for real-world deployment. 
Speech emotion recognition systems often predict a consensus value generated from the ratings of multiple annotators. However, these models have limited ability to predict the annotation of any one person. Alternatively, models can learn to predict the annotations of all annotators. Adapting such models to new annotators is difficult as new annotators must individually provide sufficient labeled training data. We propose to leverage inter-annotator similarity by using a model pre-trained on a large annotator population to identify a similar, previously seen annotator. Given a new, previously unseen, annotator and limited enrollment data, we can make predictions for a similar annotator, enabling off-the-shelf annotation of unseen data in target datasets, providing a mechanism for extremely low-cost personalization. We demonstrate our approach significantly outperforms other off-the-shelf approaches, paving the way for lightweight emotion adaptation, practical for real-world deployment.
\end{abstract}


\section{Introduction}
% General ideas (at the end describe what you are doing)
Speech Emotion Recognition (SER) is the automatic classification of emotion from speech~\cite{schuller2018speech}. SER models are typically trained on datasets with labels that are aggregated from many distinct annotators. The benefit of working with aggregated labels is that it mitigates the variability across annotators. Yet, aggregation also removes critical information about an individual annotator's perception~\cite{zhang2017predicting, han2017hard}. Emotion recognition is an inherently subjective process, and differing opinions should not necessarily be aggregated into a single opinion. Previous work has demonstrated that using the variability within individual annotators can be useful for emotion prediction, as well improving within-corpus models~\cite{10.1145/3136755.3136792,wu2023estimating,tavernor24_interspeech,zhang2017predicting}. An alternative approach is to create SER models that predict the perception of an individual annotator. Recent work has demonstrated the feasibility of such approaches~\cite{tavernor24_interspeech}. These techniques retain varied perception by modeling individual annotator labels. However, it is not clear if such models can be extended to a new dataset with new annotators without expensive model retraining.

% Present the problem in detail
Annotator-specific modeling approaches generally assume that there is sufficient data from each annotator to learn an accurate model for that annotator. When a dataset does not have sufficient annotations from individual annotators, additional data collection may be required. Even with sufficient data per annotator, finetuning such a model can be both costly and resource-intensive. We need an alternative approach: one that permits personalization while allowing for low-cost and generalizable deployment, particularly given the inherent subjectivity of the domain~\cite{steidl2005all,10832302}. Recent work has demonstrated that it is possible to accurately model a large number of sparse crowdsourced annotators~\cite{tavernor24_interspeech,listener-dependent}. However, these works only briefly address cross-corpus performance, focusing only on the acoustic adaptation of the model to new data. To our knowledge only one work has considered annotator-specific adaptation to unseen annotators, and this work only considers further finetuning on within-corpus held-out annotators~\cite{listener-dependent}. There remains an open question centered on how to leverage the similarity between the perception of different annotators for low cost generalizable cross-corpus deployment. In this work we hypothesize that annotators have inherent similarities that can be leveraged. Thus, a new \textbf{target} annotator population is likely composed of annotators who are similar to those in an existing, or \textbf{source}, population. We can leverage the existing knowledge from similar source annotators to make personalized predictions for the new target annotators. 

% Method presented to address the open questions
We present a novel lightweight cross-corpus annotator-specific modeling framework and demonstrate the efficacy of our approach across multiple well-known SER datasets. We pre-train annotator-specific models on MSP-Podcast, a large-scale emotion dataset with over 10,000 annotators~\cite{lotfian2017building}, that serves as our source dataset. Then, for each annotator in a given target dataset (i.e., MSP-Improv~\cite{improv}, IEMOCAP~\cite{iemocap}, MuSE~\cite{jaiswal2019muse}), we identify the most similar source annotator pre-trained model. Our key assumption is that given a new annotator in a target dataset (``target annotator''), there exists an annotator in the source dataset (``source annotator'') who perceives emotion similarly, independent of the specific speech segments they labeled. We assume that we have enrollment data for this new target annotator. We then assert that the ``most similar'' source annotator is the one whose pre-trained model performs most accurately, in an off-the-shelf fashion, on the target annotator's enrollment data. We use the pre-trained most similar source annotator model on the target annotator's held out test data. We compare our annotator-specific model to 1) an aggregate model trained to predict the average over all evaluations for a given sample (``aggregate prediction''), 2) the aggregate model finetuned for a single epoch (similar to having observed the enrollment data one time, as in our model), and 3) the aggregate model finetuned using all available training and validation data. The last approach is not a baseline, instead it provides an upper bound indicating what could be done given considerably more computation and labeled data. 

%This paper presents a novel approach for low-cost cross-corpus annotator-specific modeling. We test five research questions, outlined in detail in the next section.
We find that the proposed annotator-specific approach significantly outperforms aggregate prediction within-corpus, demonstrating the importance of considering annotator-specific perceptions to improve SER performance. We then explore the extension from within-corpus annotators to new unseen annotators in cross-corpus settings and find that the proposed lightweight cross-corpus adaptation method significantly outperforms other off-the-shelf annotator-specific approaches across all datasets. This holds for both individual annotator performance and when predicting the conventional aggregated label by taking an average of the output of the multiple annotator-specific models. Further, using a small enrollment set for each annotator, one that is much smaller than that annotator's full set, retains similar performance to using the full set. Finally, we demonstrate the consistency in the pairing between source and target annotators, highlighting the stability of the approach. 

This work highlights the potential of incorporating individual annotator predictions and perceptions to improve SER generalizability. Our approach is lightweight and can be deployed off-the-shelf, given a small enrollment set, making it particularly useful when capturing specific perceptual differences is necessary but full model retraining is impractical. 

\section{Research Questions}
We consider five distinct research questions that provide insight into the ability and effectiveness of our approach to identify similar annotators. 

\textbf{Research Question 1 (RQ1):} \textit{What is the relationship between annotator similarity and cross-corpus model performance?} We measure similarity between a target annotator and source annotator by deploying the pre-trained source annotator prediction head on the training data (``enrollment data'') for a given target annotator. We select the source annotator prediction head with the highest performance on the enrollment data as the most similar annotator. We hypothesize that the use of the most similar annotator will improve performance compared to a randomly selected pre-trained annotator prediction head. We investigate the relationship between the performance of the selected source annotator prediction head on a target annotator's enrollment data and the resulting performance on that annotator's test data, with poor enrollment data correlation suggesting that no similar annotator could be identified, while excellent enrollment data correlation suggests that a very similar annotator could be identified.

\textbf{Research Question 2 (RQ2):} \textit{What is the impact of source annotator prediction head accuracy with respect to performance on the target annotators?} In this case, we ask if the ``quality'' of a source annotator prediction head (i.e., how well it captures the source annotator's patterns) impacts the performance on a target annotator population. For example, the correlation between the source annotator and target annotator may be high, but the prediction head of the source annotator was not accurately learned for that source annotator. We hypothesize that source annotators who are accurately learned will perform more accurately on target annotators, compared to those who are not accurately learned. 
 %We then consider that there may be cases where the enrollment data correlation is good, however, the similar pre-trained annotator may have been poorly learned. If the original annotator was not well learned then perhaps this pre-trained annotator prediction performs artifically well on the training data by chance but does not generalize well to the test data. 

\textbf{Research Question 3 (RQ3):} \textit{What is the impact of enrollment data size on model performance?} We test the relationship between cross-corpus model performance given differing amounts of enrollment data: $N$ annotations per new annotator for $N \in [5,10,15,20,25,30]$. 

\textbf{Research Question 4 (RQ4):} \textit{How stable is the selection of a similar source annotator for each target annotator?} We evaluate the proposed method over multiple cross-validation folds and across multiple random seeds. Each iteration provides one pairing between source and target annotators. We calculate the entropy of these pairings over all folds and seeds for each annotator. We assert that if a target annotator shares a similar perception with a source annotator, we should expect that this source annotator is selected above chance for individual target annotators. 

\textbf{Research Question 5 (RQ5):} \textit{Can the prediction of individual annotators be merged to create an accurate aggregated label?} The conventional approach in SER is to predict aggregate labels. In this final research question we ask whether we can use multiple annotator-specific predictions in a cross-corpus setting to predict the aggregate label in a low-cost manner. We anticipate that this approach will outperform an off-the-shelf approach, one in which an aggregate model is learned on one dataset and applied to another. We do not anticipate that the proposed approach will outperform a model that is fine-tuned on those secondary corpora.

\section{Related Work}
Previous work has focused on evaluating perceptual differences among annotators. In some previous work, the individual annotator labels are used to create new labels or define additional multi-task targets to improve the within-corpus model performance~\cite{10.1145/3136755.3136792,wu2023estimating}. More relevantly, some work has considered the prediction of annotators on subjective tasks. For example, previous work has considered grouping similar annotators to create a new multi-task learning target based on majority and minority perception classes~\cite{10417135}. Some works have considered annotator-specific prediction ~\cite{tavernor24_interspeech,davani-etal-2022-dealing,9666407,s22145245,listener-dependent} or annotator-specific biases~\cite{9679002} in SER. Most relevantly, we expand on the multi-task frameworks of~\cite{tavernor24_interspeech,davani-etal-2022-dealing}, where a separate classification head is used to make predictions for each annotator. This previous work has shown accurate performance across crowdsourced annotators within-corpus and cross-corpus. However, the cross-corpus experiments have been limited, focusing only on using all pre-trained annotators to make predictions rather than individual annotator predictions. We seek to use information provided by annotator-specific predictions to improve cross-corpus performance without training the model. 

To address generalizability in SER, prior work has investigated the importance of learning similar hidden representations for emotion expression across datasets~\cite{gideon2019improving,abdelwahab2018domain}. An alternative previous approach has been to augment or generate new data to help models be more generalizable~\cite{9687987,zhang2011unsupervised}. Ando et al. compare the performance of individual annotator models using finetuning (one model per annotator), auxiliary input (where annotator embeddings are input to an adapter layer), and sub-layer weighting (where each annotator has individual weights for later model layers)~\cite{listener-dependent}. The authors investigate both finetuning the full model and finetuning only an annotator embedding that is input to the model on adaptation data. The authors demonstrate the effectiveness of this adaptation for categorical emotion on within-corpus annotator adaptation. However, our focus is on deployment without any model training. Instead of finetuning on within-corpus held-out annotators, we investigate the feasibility of off-the-shelf deployment to completely unseen data. 

\section{Datasets}
\subsection{Data Description}
\label{sec:data}
\textbf{MSP-Podcast} is a dataset of non-acted speech taken from podcasts~\cite{lotfian2017building}. We use this dataset for model pre-training. We use release 1.11, which contains 12,913 different annotators in the training set. We pre-train on MSP-Podcast, as it is the largest and most naturalistic dataset, and because its large number of annotators increases the likelihood of finding similar annotators when comparing to target datasets~\cite{chen23b_interspeech}.% These 12,913 are the prediction targets for our pre-trained annotator-specific model. 

We evaluate cross-corpus performance on three different SER datasets. We use \textbf{MSP-Improv} and \textbf{IEMOCAP}, which are acted datasets using both scripted and improvised sessions of dialogue between two speakers~\cite{improv, iemocap}. MSP-Improv contains 1,496 different annotators, while IEMOCAP only contains six annotators. The final dataset used for adaptation is \textbf{MuSE}~\cite{jaiswal2019muse}. MuSE consists of monologue speech in response to emotional stimuli of 28 college students, with sessions recorded in both stressed and unstressed conditions. We use the out-of-context label annotations (utterances annotated in random order) from 160 annotators.

For each dataset, during both pre-training and adaptation, we first remove all annotators \camerareadychanges{from the training set }{}that have not annotated at least 30 samples in the dataset to enable an accurate measurement of correlation~\cite{ae7889f8-f8e6-35c0-919b-64c18f0b9197,lawrence1992assay}. We then remove all annotators from the validation and test set who do not appear in the training sets. After processing, the MSP-Podcast dataset consists of 1,998 annotators\camerareadychanges{}{ and 134,088 samples}. The exact number of annotators on the adaptation datasets varies for MSP-Improv and MuSE because the training set varies during cross-validation. There are $513.4$±$47.399$ and $70.8$±$2.713$ annotators on MSP-Improv and MuSE respectively. IEMOCAP has a small number of annotators - the same four annotators are present in all folds. \camerareadychanges{}{MSP-Improv, MuSE, and IEMOCAP have 8,438, 2,647 and 9,999 samples respectively.}

We focus on dimensional emotion recognition, predicting \textbf{activation} (energy) and \textbf{valence} (positivity) from speech.

% MuSE and IEMOCAP both contain self-report labels. However, we focus only on MuSE for self-report experiments as IEMOCAP self-report labels are not annotated at the time of speech recording. Instead, they are labeled afterwards, with the speakers listening back to their own speech, which we believe makes their labels less akin to realistic self-report annotations. The self-annotations of MuSE are taken at the end of monologue sessions, so utterances within the same monologue all contain the same self-report label. 

\subsection{Dataset Preprocessing}
\label{sec:preprocess}
We \textbf{preprocess} labels for all datasets. We form two label sets: a) the individual annotator label and b) the aggregated label. The individual annotator label describes how an individual annotator evaluated a given piece of data (e.g., $y_{val,i}$ for the valence annotation for annotator $i$). The aggregate label is the average over all annotators who have evaluated the same piece of data (e.g., $\bar{y}_{val}$ for valence). We transform both label sets using min-max scaling into a $[-1,1]$ range. The scaling parameters are determined over the entire annotator population and are not annotator-specific.

\camerareadychanges{We perform five-fold cross-validation for all cross-corpus experiments, where the folds are defined for each of the three target datasets}{For each of the three target datasets, we perform five-fold cross-validation generating adaptation (training) and testing folds}. For IEMOCAP and MSP-Improv we create session-independent folds to ensure speaker-independence and remove any risk of potential crosstalk. For MuSE we randomly generate speaker-independent validation folds\footnote{Our code and information on cross-validation folds is available on our GitHub https://github.com/chailab-umich/MoreSimilarThanDissimilar}.

\section{Methods}
\subsection{Model Architecture}
The input to the model is frozen BERT~\cite{devlin-etal-2019-bert} CLS embeddings, as has previously been used for this task~\cite{tavernor24_interspeech}, and frozen WavLM~\cite{9814838} embeddings. Previous work used Wav2Vec2~\cite{baevski2020wav2vec} embeddings~\cite{tavernor24_interspeech}). WavLM has demonstrated relative performance improvement compared to Wav2Vec2 on SER~\cite{feng2023peft}. We concatenate the mean-pooled WavLM embedding and BERT CLS token after applying dropout. The concatenated embedding then goes through a linear layer. It is then passed through two linear layers for activation and a separate two linear layers for valence to get activation and valence embeddings. Finally, the activation and valence embeddings each pass through a final linear prediction layer. All layers before the prediction layer are size 256 and use ReLU activation. 

The model predictions are either at the individual annotator level or at the aggregate level. The \textbf{individual annotator model} takes in an input speech sample and makes a prediction, $\hat{p}_{val,i}$, for each annotator, $i$, who annotated the sample. The prediction target is $p_{val,i}$, that annotator's annotation of valence. The \textbf{aggregated baseline model} takes in the same input sample and makes a single prediction $\hat{y}_{val}$. The prediction target is $y_{val} \equiv \frac{\sum_i p_{val,i}}{N}$ (for $N$ annotations on the sample), the average of all of the valence annotations for the input sample. The same definitions hold for activation. The individual annotator models have 1,998 prediction heads for each of activation and valence to predict each annotator (Section~\ref{sec:data}). The aggregate model has one prediction layer for each of activation and valence. 

\subsection{Within-Corpus Training}
\label{sec:pretrain}
We train the annotator-specific models on MSP-Podcast. The loss function is Lin's Concordance Correlation Coefficient (CCC)~\cite{lawrence1989concordance} loss, in line with previous works~\cite{tavernor24_interspeech,wagner2023dawn}. In the individual annotator model, the loss function captures the difference between the individual predictions (e.g., $\hat{p}_{val,i}$) and the known annotations of the individual annotators (e.g., $p_{val,i}$) over all annotators in the batch, as in~\cite{tavernor24_interspeech}. We use a multitask objective, equally weighting activation and valence~\cite{tavernor24_interspeech,sampath2025efficient}. In the aggregate model, we equally weight valence and activation CCC (e.g., comparing $\hat{y}_{act}$ and $y_{act}$ for each sample in the batch). The annotator-specific prediction heads are trained over six seeds. In each seed, we initialize the prediction heads by training an aggregate prediction for five epochs. 

\subsection{Cross-Corpus Ind. Annotator Mapped (IA PT-Mapped)}
\label{sec:iamap}
The model with annotator-specific prediction heads trained on MSP-Podcast (Section~\ref{sec:pretrain}), is used, off-the-shelf, to predict the emotion perception of every sample of a given target annotator's enrollment data (the training data evaluated by a given annotator) to identify the prediction head associated with the pre-trained annotator that is most well-aligned with each target annotator over the three datasets. This is the proposed \textbf{IA PT-Mapped} model. 

We identify similar source-target annotator pairs by identifying the source annotator prediction head that performs best on a target's enrollment data based on CCC (metrics are discussed in more detail in Section~\ref{sec:metrics}). We use this source annotator prediction head, without adaptation, to make predictions for the target annotator on the test partition of the dataset, and use only the target annotator labels to determine similarity. 

The research questions analyze both the ability to predict the annotations of individual annotators and the ability to predict an aggregate label. The aggregate ground truth is an average of the original annotations for that sample.  The IA PT-Mapped aggregate prediction is the average of the outputs of the chosen source annotator prediction heads.
%RQ5 necessitates the creation of an aggregate prediction from the individual predictions. We create an aggregate prediction following the definition of an aggregate ground truth, which is the average of the annotations provided by the individuals who annotated a given sample. We mirror this at the sample-level by deploying the selected source annotator prediction heads and averaging the predictions.% for each of the target annotators who evaluated a given sample. 

\subsection{Cross-Corpus Individual Annotator Baselines}
\label{sec:ia_base}
\noindent\textbf{Individual Random Map (IA PT-Random)}:
% \label{sec:iapt}
%This baseline will indicate the importance of identifying the most similar source annotator to a given target annotator (RQ1). 
We randomly assign each target annotator a source annotator prediction head. We create individual and aggregate predictions as in Section~\ref{sec:iamap}.%The pre-trained model (1,998 annotator-specific prediction heads) trained on MSP-Podcast is used, off-the-shelf, to predict the emotion perception of every sample in the target training set. In this case, we randomly assign each target annotator a source annotator model. We create aggregate predictions as described in Section~\ref{sec:iamap}.

\noindent\textbf{Individual Pre-trained (IA PT and IA PT-All)}:
% \label{sec:iapt}
We evaluate \textbf{IA PT} within corpus. At the sample-level, we deploy the prediction heads associated with the subset of annotators who evaluated the given sample. Cross-corpus, we cannot deploy the model in this manner. The mapped and random models provide a method to select pre-trained source annotator prediction heads for a specific annotator population. However, this may not always be possible if a target dataset does not have training data available. Instead of making predictions using source annotators chosen for the target annotators of a given sample, we will use all 1,998 annotator-specific prediction heads. We then average the 1,998 estimates for each sample and assign this average as the prediction for each target annotator. We refer to this approach as \textbf{IA PT-All}.

\subsection{Cross-Corpus Aggregate Baseline}
\label{sec:agg_base}
\noindent\textbf{Aggregate Baseline Pre-trained (Agg. PT)}: 
The model described in Section~\ref{sec:pretrain} is used in an off-the-shelf manner to predict the aggregated labels in the target dataset.

\noindent\textbf{Aggregate Baseline Full (Agg. FT-Full)}: 
We finetune the Agg. PT model using the same method as described in Section~\ref{sec:pretrain} on the target dataset's training data until early stopping triggers on the target dataset's validation data using patience of 10. This baseline is an upper bound to our proposed non-finetuning approach as we allow the model to train as many epochs as required on the full training set. Further, the model is provided with additional data that our proposed method does not use in the form of the validation data. We present these results to provide context for the results of the lightweight off-the-shelf proposed approach.

\noindent\textbf{Aggregate Baseline Finetuned, One Epoch (Agg. FT-1)}: 
We finetune the Agg. PT model in the same way as Agg. FT-Full on the target dataset's training data, but this time for one epoch only. The lightweight adaptation evaluates the training data once and this allows us to ask how a finetuning approach would do if offered a single chance to adapt with these data. 

\noindent\textbf{Aggregate Baseline Ground Truth (Agg. Ground Truth)}: The final baseline is an oracle method. We calculate the aggregate label over all annotators who annotated a given sample. We then assign this label to each annotator. If the Agg. Ground Truth model is effective, it suggests that individual annotators are well captured by an aggregate label. As such, this baseline allows us to understand when other baseline methods, particularly Agg. FT-Full, are likely to be effective. 

\section{Metrics}
\label{sec:metrics}

The first metric is focused on the performance of a model prediction head for a single annotator. We call this metric $CCC_{ind}$. We create two vectors for each annotator, their annotations and the model's prediction. We calculate the CCC between the two. We average this value over all annotators who annotated at least two samples.

The second metric is focused on the conventional aggregate annotations (the average of all annotators who annotated a given sample). We call this metric $CCC_{agg}$. We create two vectors over all samples in a dataset, the ground truth aggregate annotations and a given model's predictions. We calculate the CCC between the two. 

\section{Results and Discussion}

For all applicable results we report significance using a paired t-test at a 95\% confidence. 

\subsection{Within-Corpus Performance}
We first report the within-corpus $CCC_{agg}$ performance of the models described in Section~\ref{sec:pretrain} on MSP-Podcast. This provides a demonstration that, within-corpus, both models are able to accurately predict the aggregated annotations. We find that considering the individual annotator perceptions (IA PT) significantly outperforms the methods without annotator-specific predictions except for $CCC_{ind}$ on valence, where performance is not significantly different (see Table~\ref{tab:pretrain}).

\begin{table}[t]
%\caption{Pre-trained model $CCC_{agg}$ on MSP-Podcast. IA-PT refers to the Individual Pre-trained Baseline and Agg. PT to the Aggregate Baseline Full (Section~\ref{sec:baseline}). \textbf{IA PT} makes the aggregate prediction using only the annotators present on the testing sample. We additionally report results where all annotators are used to make all test predictions, as would be the case in a default unseen dataset (\textbf{IA PT-All}).}
\caption{Within-corpus CCC on MSP-Podcast. IA-PT, IA-PT-All are described in Section~\ref{sec:ia_base} and Agg. PT is described in (Section~\ref{sec:agg_base}). .  The best performances are bolded for each column, $*$ indicates a significant decrease in performance compared to IA-PT. }
\vspace{-10pt}
\label{tab:pretrain}
\begin{center}
\begin{tabular}{ccccc}
\hline
\textbf{Model} & \multicolumn{2}{c}{\textbf{Activation}} & \multicolumn{2}{c}{\textbf{Valence}} \\ 
& $CCC_{agg}$ & $CCC_{ind}$ & $CCC_{agg}$ & $CCC_{ind}$ \\ \hline
%\textbf{\begin{tabular}[c]{@{}c@{}}IA PT\end{tabular}} &0.658$\pm$0.003 & 0.393$\pm$0.003 & 0.647$\pm$0.002 & 0.398$\pm$0.002 \\
\textbf{\begin{tabular}[c]{@{}c@{}}IA PT\end{tabular}} &\textbf{.658$\pm$.003} & \textbf{0.393$\pm$.003} & \textbf{.647$\pm$.002} & .398$\pm$.002 \\
%\textbf{\begin{tabular}[c]{@{}c@{}}IA PT*\end{tabular}} &0.588$\pm$0.007↓$\dagger$ & 0.389$\pm$0.002↓$\dagger$ & 0.578$\pm$0.002↓$\dagger$ & 0.393$\pm$0.002↓$\dagger$ \\
\textbf{\begin{tabular}[c]{@{}c@{}}IA PT-All\end{tabular}} &.588$\pm$.007* & .389$\pm$.002* & .578$\pm$.002* & .393$\pm$.002* \\
%\textbf{\begin{tabular}[c]{@{}c@{}}Agg. PT\end{tabular}} &0.599$\pm$0.004↓$\dagger$ & 0.375$\pm$0.003↓$\dagger$ & 0.587$\pm$0.002↓$\dagger$ & 0.399$\pm$0.005↑ \\ \hline
\textbf{\begin{tabular}[c]{@{}c@{}}Agg. PT\end{tabular}} &.599$\pm$.004* & .375$\pm$.003* & .587$\pm$.002* & \textbf{.399$\pm$.005} \\ \hline
\end{tabular}
\end{center}
\vspace{-15pt}
\end{table}

The first within-corpus result is the Individual Pre-trained (IA PT) model. This model generates predictions for each of the annotators for a given sample and then averages the result. The average is compared to the aggregate ground truth. The performance for valence and activation is comparable, at 0.658$\pm$0.003 and 0.647$\pm$0.002, respectively. The IA PT-All model does the same, except in this case it makes a prediction for each of the 1,998 prediction heads. This is the default in an unseen dataset. The results are statistically significantly lower, at 0.588$\pm$0.007 and 0.578$\pm$0.002, respectively. The Aggregate Baseline (Agg. PT) predicts the aggregate directly. The results are comparable to the IA PT-All model. The results highlight the benefit of aligning the prediction heads with the annotators who annotated a given sample (further discussed in RQ1).

\subsection{Cross-Corpus Performance}
\label{sec:thirdperson}
We first evaluate $CCC_{ind}$, which captures the ability of a model to predict annotators' annotations on unseen data. We observe that the IA PT-Mapped model (Section~\ref{sec:iamap}) significantly outperforms all other approaches, except two cases (Table~\ref{tab:all}, valence on IEMOCAP and MSP-Improv).

\begin{table*}[t]
%\caption{$CCC$ Results. IA is the Individual Annotator model, Agg. is the Aggregate Baseline, Mapped indicates predictions were mapped to similar annotators, PT indicates only pre-training, FT-1 indicates 1 epoch of finetuning, and FT-Full indicates finetuning until early stopping. $\dagger$ and $*$ indicate significant decrease or increase respectively compared to the proposed approach. Best results are highlighted in bold, if the best results is FT-Full, we also highlight the second best result, as this model is an upper-bound model.}
\caption{Cross-corpus $CCC$ results. IA are Individual Annotator models (Section~\ref{sec:iamap}, ~\ref{sec:ia_base}), Agg. are Aggregate Baselines (Section~\ref{sec:agg_base}), PT is only pre-training, FT-1 is 1 epoch of finetuning, and FT-Full is finetuning until early stopping. $\dagger$ and $*$ indicate significant decrease or increase respectively compared to the proposed IA PT-Mapped approach. Best results over the non-oracle methods are highlighted in bold. }
\begin{center}
% \addtolength{\tabcolsep}{-0.4em}
\begin{tabular}{lllllllll}
\hline
&\multirow{2}{4em}{\textbf{Model}} & \multirow{2}{4em}{\textbf{Training}} & \multicolumn{3}{c}{\textbf{Activation}} & \multicolumn{3}{c}{\textbf{Valence}} \\ 
& & & \multicolumn{1}{c}{\textbf{IEMOCAP}} & \multicolumn{1}{c}{\textbf{MSP-Improv}} & \multicolumn{1}{c}{\textbf{MuSE}} & \multicolumn{1}{c}{\textbf{IEMOCAP}} & \multicolumn{1}{c}{\textbf{MSP-Improv}} & \multicolumn{1}{c}{\textbf{MuSE}} \\

\hline
&&&\multicolumn{6}{c}{\textbf{$CCC_{ind}$}}\\
\hline

\multirow{4}{4em}{Off-the-shelf}& IA & PT-Mapped & \textbf{0.520±0.030} & \textbf{0.415±0.062} & \textbf{0.367±0.057} & 0.452±0.044 & 0.329±0.029 & \textbf{0.458±0.028} \\
%&IA & PT-Random & 0.425±0.041↓$\dagger$ & 0.287±0.051↓$\dagger$ & 0.211±0.042↓$\dagger$ & 0.315±0.086↓$\dagger$ & 0.235±0.024↓$\dagger$ & 0.283±0.033↓$\dagger$ \\
%&IA & PT* & 0.473±0.034↓$\dagger$ & 0.369±0.060↓$\dagger$ & 0.293±0.049↓$\dagger$ & 0.429±0.060↓$\dagger$ & 0.333±0.028↑* & 0.446±0.029↓$\dagger$ \\
%& Agg. & PT & 0.475±0.029↓$\dagger$ & 0.320±0.054↓$\dagger$ & 0.258±0.045↓$\dagger$ & 0.418±0.049↓$\dagger$ & 0.290±0.022↓$\dagger$ & 0.391±0.026↓$\dagger$ \\
&IA & PT-Random & 0.425±0.041$\dagger$ & 0.287±0.051$\dagger$ & 0.211±0.042$\dagger$ & 0.315±0.086$\dagger$ & 0.235±0.024$\dagger$ & 0.283±0.033$\dagger$ \\
&IA & PT* & 0.473±0.034$\dagger$ & 0.369±0.060$\dagger$ & 0.293±0.049$\dagger$ & 0.429±0.060$\dagger$ & 0.333±0.028* & 0.446±0.029$\dagger$ \\
& Agg. & PT & 0.475±0.029$\dagger$ & 0.320±0.054$\dagger$ & 0.258±0.045$\dagger$ & 0.418±0.049$\dagger$ & 0.290±0.022$\dagger$ & 0.391±0.026$\dagger$ \\
\hline
%\multirow{2}{4em}{Finetuned} &Agg. & FT-1 & 0.476±0.034↓$\dagger$ & 0.343±0.062↓$\dagger$ & 0.251±0.045↓$\dagger$ & 0.489±0.057↑* & 0.358±0.045↑* & 0.403±0.029↓$\dagger$ \\
%&Agg. & FT-Full & 0.489±0.032↓$\dagger$ & 0.360±0.062↓$\dagger$ & 0.248±0.048↓$\dagger$ & \textbf{0.569±0.074↑*} & \textbf{0.397±0.043↑*} & 0.410±0.029↓$\dagger$ \\
\multirow{2}{4em}{Finetuned} &Agg. & FT-1 & 0.476±0.034$\dagger$ & 0.343±0.062$\dagger$ & 0.251±0.045$\dagger$ & 0.489±0.057* & 0.358±0.045* & 0.403±0.029$\dagger$ \\
&Agg. & FT-Full & 0.489±0.032$\dagger$ & 0.360±0.062$\dagger$ & 0.248±0.048$\dagger$ & \textbf{0.569±0.074*} & \textbf{0.397±0.043*} & 0.410±0.029$\dagger$ \\
%\hline\multicolumn{2}{c}{\textbf{Agg. Ground Truth}} & \textbf{0.729±0.020↑*} & \textbf{0.517±0.045↑*} & 0.352±0.054↓$\dagger$ & \textbf{0.894±0.022↑*} & \textbf{0.754±0.026↑*} & \textbf{0.602±0.014↑*} \\
\hline
%Oracle & Agg. & Ground Truth & 0.729±0.020↑* & 0.517±0.045↑* & 0.352±0.054↓$\dagger$ & 0.894±0.022↑* & 0.754±0.026↑* & 0.602±0.014↑* \\
Oracle & Agg. & Ground Truth & 0.729±0.020* & 0.517±0.045* & 0.352±0.054$\dagger$ & 0.894±0.022* & 0.754±0.026* & 0.602±0.014* \\

\hline
\\
&&&\multicolumn{6}{c}{\textbf{$CCC_{agg}$}}\\
\hline

\multirow{4}{4em}{Off-the-shelf}& IA & PT-Mapped & 0.631±0.030 & \textbf{0.610±0.069} & \textbf{0.719±0.079} & 0.516±0.041 & 0.538±0.050 & 0.625±0.045 \\
%& IA & PT-Random & 0.585±0.039↓$\dagger$ & 0.438±0.071↓$\dagger$ & 0.560±0.117↓$\dagger$ & 0.406±0.099↓$\dagger$ & 0.434±0.056↓$\dagger$ & 0.561±0.063↓$\dagger$ \\
%& IA & PT* & 0.601±0.027↓$\dagger$ & 0.548±0.075↓$\dagger$ & 0.620±0.111↓$\dagger$ & 0.504±0.063↓ & 0.517±0.051↓$\dagger$ & 0.611±0.048↓$\dagger$ \\
%& Agg. & PT & 0.629±0.019↓ & 0.479±0.066↓$\dagger$ & 0.619±0.101↓$\dagger$ & 0.505±0.048↓$\dagger$ & 0.466±0.050↓$\dagger$ & 0.544±0.070↓$\dagger$ \\
& IA & PT-Random & 0.585±0.039$\dagger$ & 0.438±0.071$\dagger$ & 0.560±0.117$\dagger$ & 0.406±0.099$\dagger$ & 0.434±0.056$\dagger$ & 0.561±0.063$\dagger$ \\
& IA & PT* & 0.601±0.027$\dagger$ & 0.548±0.075$\dagger$ & 0.620±0.111$\dagger$ & 0.504±0.063 & 0.517±0.051$\dagger$ & 0.611±0.048$\dagger$ \\
& Agg. & PT & 0.629±0.019 & 0.479±0.066$\dagger$ & 0.619±0.101$\dagger$ & 0.505±0.048$\dagger$ & 0.466±0.050$\dagger$ & 0.544±0.070$\dagger$ \\
\hline
%\multirow{2}{4em}{Finetuned} & Agg. & FT-1 & 0.664±0.037↑* & 0.549±0.093↓$\dagger$ & 0.627±0.092↓$\dagger$ & 0.584±0.034↑* & 0.548±0.067↑ & 0.638±0.037↑* \\
%& Agg. & FT-Full & \textbf{0.695±0.015↑*} & 0.586±0.078↓$\dagger$ & 0.637±0.090↓$\dagger$ & \textbf{0.678±0.045↑*} & \textbf{0.597±0.055↑*} & \textbf{0.655±0.039↑*} \\
\multirow{2}{4em}{Finetuned} & Agg. & FT-1 & 0.664±0.037* & 0.549±0.093$\dagger$ & 0.627±0.092$\dagger$ & 0.584±0.034* & 0.548±0.067 & 0.638±0.037* \\
& Agg. & FT-Full & \textbf{0.695±0.015*} & 0.586±0.078$\dagger$ & 0.637±0.090$\dagger$ & \textbf{0.678±0.045*} & \textbf{0.597±0.055*} & \textbf{0.655±0.039*} \\
\hline
\end{tabular}
\label{tab:all}
\vspace{-15pt}
\end{center}
\end{table*}

The oracle Agg. Ground Truth model outperforms all other approaches, except for activation on MuSE, where IA PT-Mapped performs most accurately. The oracle performance shows that the aggregate label is well correlated with annotators and points to the relevance of the three aggregate models (Agg. PT, Agg. FT-1, and Agg. FT-Full) as baselines. 

In valence classification for IEMOCAP and MSP-Improv, the finetuned aggregate models improve performance over all of the off-the-shelf models, including IA PT-Mapped. We anticipate that this is due to the strong relationship between the aggregate label and any single individual (Table~\ref{tab:all}) and the relationship between valence and text~\cite{wagner2023dawn,calvo2010affect}. Further, IEMOCAP and MSP-Improv are both scripted and the cross-validation folds are session-dependent, resulting in consistent text-emotion pairs over the folds. The model is likely learning to memorize the words associated with emotional valence.

\subsection{Research Questions}

As IEMOCAP only consists of a very small number of annotators, we focus our analyses on MuSE and MSP-Improv. 

\subsubsection{Research Question 1 (RQ1): What is the relationship between annotator similarity and cross-corpus model performance?} 

\begin{figure}[t]
\centerline{\includegraphics[width=0.5\textwidth]{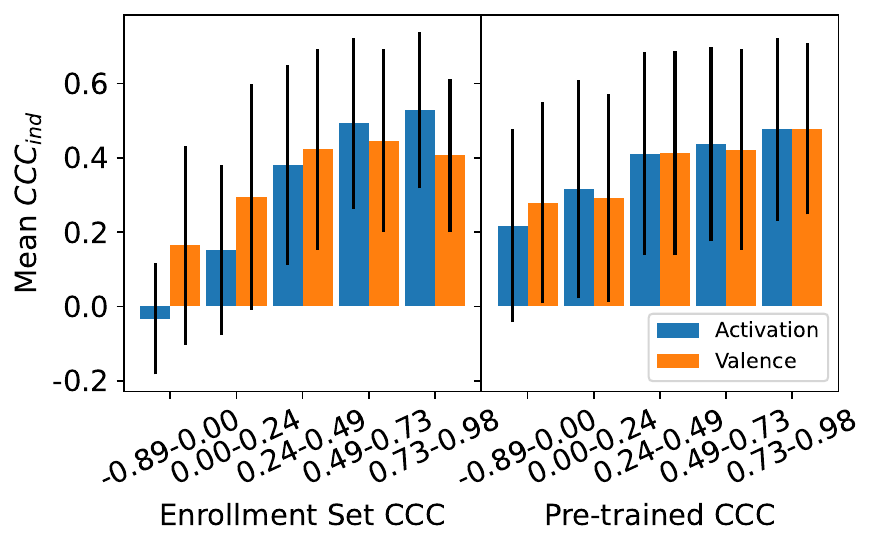}}
\caption{$CCC_{ind}$ performance over MuSE and MSP-Improv for all seeds and folds, grouped by enrollment $CCC$. The first bin is $CCC < 0$ all other bins are equal width.}
\label{fig:enroll}
\vspace{-15pt}
\end{figure}

We consider the best performing individual annotator model (IA PT-Mapped) on MSP-Improv and MuSE (IEMOCAP has only four consistent annotators).  We ask whether the performance on enrollment data is indicative of performance on test data.  We first calculate the $CCC_{ind}$ over the enrollment data for each source-target pair that occurs throughout any of the folds and random seeds.  We repeat this for the test data.  We first present a histogram showing the relationship between the two values (Figure~\ref{fig:enroll}, left), which suggests that as CCC on enrollment data increases, performance on the test data increases.  Next, we calculate the Pearson's Correlation Coefficient (PCC) between both resulting vectors (Table~\ref{tab:enroll}).  We observe that there appears to be a relationship between this PCC value and the overall performance of the model (compare MSP-Improv and MuSE, valence and activation in Table~\ref{tab:all}).  However, we caution against overly strong conclusions given the focus on only two datasets.

%We calculate the CCC between the annotator's enrollment set $CCC_{ind}$ with their test set $CCC_{ind}$ for each dataset across all seeds and folds (Table~\ref{tab:enroll}). This correlation confirms the previous conclusions. There is notable correlation with the Agg. Ground Truth, suggesting in part that good performance on the enrollment set corresponds to more similarity to the aggregate label. However, in all cases (except Activation on MuSE), the relationship between the enrollment data performance and final test performance is much more important for the IA PT-Mapped model as expected. We find on the completely unscripted dataset (MuSE) that the enrollment set performance effect on valence becomes much more prominent compared to the other datasets.

\begin{table}[t]
%\caption{Correlation of each annotator's enrollment set $CCC_{ind}$ with their test set $CCC_{ind}$ for each dataset across all seeds and folds. IEMO. is the IEMOCAP dataset and Imp. is the MSP-Improv dataset.}
%\vspace{-12pt}
\caption{PCC between each annotator pair's enrollment set $CCC_{ind}$ and test set $CCC_{ind}$ over all seeds and folds }
\vspace{-10pt}
\label{tab:enroll}
\begin{center}
\begin{tabular}{ccc}
\hline
%\textbf{Dataset} & \multicolumn{2}{c}{\textbf{IA PT-Mapped}} & \multicolumn{2}{c}{\textbf{Agg. Ground Truth}} \\ 
%& Activation & Valence & Activation & Valence \\ \hline
%IEMO. & 0.617±0.198 & 0.077±0.073 & -0.095±0.065 & 0.001±0.008 \\
%Imp. & 0.461±0.070 & 0.272±0.038 & 0.426±0.063 & 0.084±0.017 \\
%MuSE & 0.239±0.084 & 0.523±0.064 & 0.254±0.068 & 0.438±0.037 \\ \hline
\textbf{Dataset} & \textbf{Activation} & \textbf{Valence} \\ \hline
%IEMOCAP & 0.617±0.198 & 0.077±0.073 \\
MSP-Improv & 0.495 & 0.321 \\
MuSE & 0.267 & 0.540 \\ \hline
\end{tabular}
\end{center}
%\vspace{-20pt}
\vspace{-15pt}
\end{table}

\subsubsection{Research Question 2 (RQ2): What is the impact of source annotator prediction head accuracy with respect to performance on the target annotators?} 

We again restrict our analysis to  MSP-Improv and MuSE. We analyze the relationship between the $CCC_{ind}$ of the selected source annotator prediction head on the target annotator's test data and the $CCC_{ind}$ observed for that source annotator during model training on the source dataset. We observe only a slight increase in performance when the source annotator was well learned for both valence and activation (Figure~\ref{fig:enroll}, right). This may be because poorly learned source annotators were not selected. For example, the median $CCC_{ind}$ on the pre-training dataset (MSP-Podcast) of the selected source annotators for MSP-Improv and MuSE (considered together) was $0.457$ and $0.384$ for activation and valence, respectively (the three right-most sets of bars in Figure~\ref{fig:enroll}).
%Therefore, the range of $CCC_{ind}$ considered is relatively small. We believe that this is because it is unlikely that a poorly learned annotator would be selected as the most similar for a given target annotator.

\subsubsection{Research Question 3 (RQ3): What is the impact of enrollment data size on model performance?} 

\begin{figure}[t]
\centering
\includegraphics[width=\linewidth]{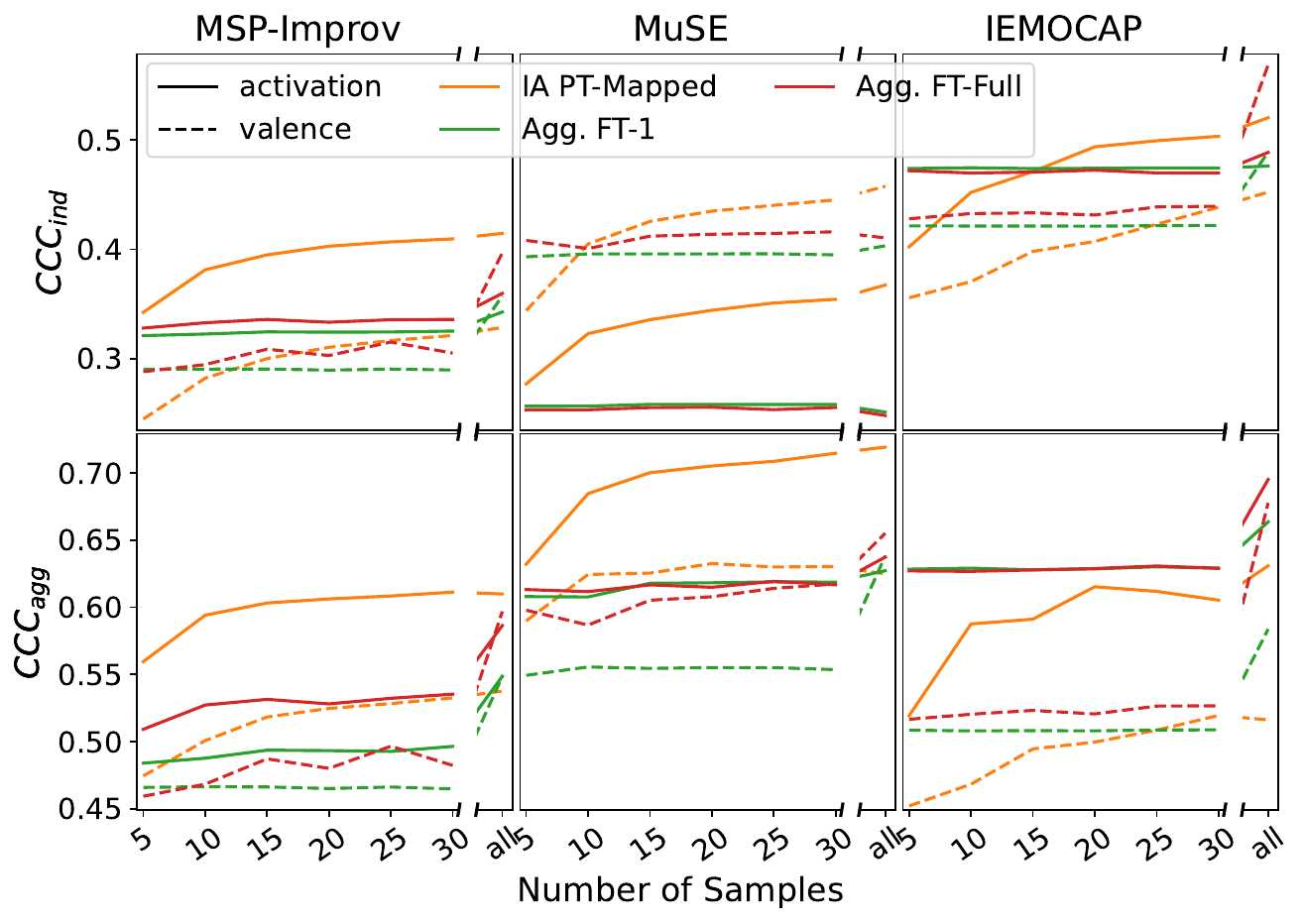}
% \vspace{-20pt}
\caption{Performance trends on each of the three datasets for $CCC_{ind}$ and $CCC_{agg}$, x-axis:the number of enrollment samples, for 5 to 30 in steps of 5, then all. The average enrollment set size for 'all' is $41.5\pm8.2$, $103.6\pm13.7$, and $4997.8\pm609.7$ for MSP-Improv, MuSE, and IEMOCAP respectively.}
\label{fig:iemo}
\vspace{-15pt}
\end{figure}

We find that less than 30 samples per annotator are required to achieve performance comparable to the entire training set for IA PT-Mapped. Most approaches begin to reach similar performance with $\sim$15-20 annotations per annotator, much less than the full set of annotations. In some cases, for both $CCC_{ind}$ and $CCC_{agg}$ (MSP-Improv activation, MuSE activation, valence) using such a small set outperforms Agg. FT-1 even when providing the entire training set to Agg. FT-1, except for MuSE valence, Agg. FT-Full is also out performed in these same cases (see Figure~\ref{fig:iemo}).

% \begin{figure*}[t]
% \centering
% \includegraphics[width=\linewidth]{figures/CCC--agg--by-Enrollment-Set-Size.pdf}
% % \vspace{-20pt}
% \caption{Performance trends on each of the three datasets for $CCC_{agg}$, x-axis: the number of enrollment samples, for 5 to 30 in steps of 5, then all.}
% \label{fig:iemo}
% % \vspace{-15pt}
% \end{figure*}

\subsubsection{Research Question 4 (RQ4): How stable is the selection of a similar source annotator for each target annotator?} 
We measure stability by calculating the entropy over the selection of similar annotators.  In every run, each target annotator selects one source annotator from the 1,998 MSP-Podcast annotators. We observe that the most unstable selection would involve choosing a different annotator at each selection point.  We consider the six random seeds and five folds (a total of thirty repetitions). We calculate entropy using $\log_2$.  Therefore, the most unstable selection results in an entropy of $4.906$. Considering only the new annotators that occur in all thirty repetitions, we find that the entropy of similar annotator selection is considerably lower: $2.478$±$0.736$, $2.305$±$0.319$, and $2.741$±$0.702$ for MSP-Improv, IEMOCAP, and MuSE respectively.  This highlights that even over different folds and seeds, we generally see the same source annotators selected.

\subsubsection{Research Question 5 (RQ5): Can the prediction of target annotators be merged to create an accurate aggregated label?} 

We evaluate $CCC_{agg}$, which captures the ability of a model to predict an aggregate label. We average the sample-level predictions for all IA models (Sections~\ref{sec:iamap} and~\ref{sec:ia_base}). We observe that IA PT-Mapped outperforms all off-the-shelf approaches across all datasets (including the off-the-shelf deployment of Agg. PT). It is generally outperformed by models finetuned on the target datasets.
%IA PT-Mapped statistically outperforms Agg. FT-1 and Agg. FT-Full on activation (MSP-Improv, MuSE) and valence (MSP-Improv). 
However, we remind that the off-the-shelf approaches do not finetune. 
It is therefore not surprising that the approaches finetuned on the aggregate labels of the target datasets outperfrom IA PT-Mapped at the same task.

\section{Conclusion}
In this paper, we have demonstrated a novel adaptation method considering the correlation between pre-trained annotators and new annotators from the new unseen data. We have shown that by using a mapping method, an untrained annotator-specific method can outperform even trained aggregate models in many cases. Furthermore, we have shown that this method is effective even when using less than 30 annotations per new annotator, enabling adaptation to new annotator perceptions with very limited labeling required of the new annotators. Future work will include investigations into finetuning of the annotator-specific models (IA PT-Mapped), providing a new avenue for personalization.

\section*{Acknowledgment}
This material is based in part upon work supported by the National Science Foundation (NSF IIS-RI 2230172) and National Institutes of Health (NIH R01MH130411).
\clearpage

\bibliographystyle{IEEEtran}
\bibliography{mybib}

\end{document}